\documentclass[aps,prl,twocolumn,showpacs,superscriptaddress]{revtex4-1}  
\usepackage{float}
\usepackage{bbm}
\usepackage{epsfig}
\usepackage{epstopdf}
\usepackage{graphicx}
\usepackage{amsmath,amssymb}
\usepackage{amsmath,bm}
\usepackage{physics}
\usepackage{color}
\usepackage{hyperref}
\usepackage{lineno,blindtext}
\usepackage[caption=false]{subfig}
\usepackage{siunitx, booktabs}
\usepackage{diagbox, eqparbox, hhline}
\usepackage{soul}
\usepackage{xcolor}
\usepackage[normalem]{ulem}
\newcolumntype{P}[1]{>{\centering\arraybackslash}p{#1}}

\begin{document}

\title{{Non-Hermitian skin effect in magnetic systems }}

\author{Kuangyin Deng}
\email{dengku@bc.edu}
\author{Benedetta Flebus}
\email{flebus@bc.edu}
\affiliation{Department of Physics, Boston College, 140 Commonwealth Avenue, Chestnut Hill, Massachusetts 02467, USA}

\begin{abstract}

Far from being limited to a trivial generalization of their Hermitian counterparts, non-Hermitian topological phases have gained widespread interest due to their unique properties. One of the most striking  non-Hermitian phenomena  is the skin effect, i.e., the localization of a macroscopic fraction of bulk eigenstates at a boundary, which underlies the breakdown of the bulk-edge correspondence. Here we develop a generic phenomenological approach to describing magnetic dissipation within a lattice model and we introduce an ``effective area law" to investigate the emergence of the skin effect in magnetic  systems. 
As a testbed of our approach, we focus on a spin-orbit-coupled van der Waals (vdW) ferromagnet with spin-nonconserving magnon-phonon interactions, finding that the magnetic skin effect emerges in an appropriate temperature regime. Our results suggest that the interference between  Dzyaloshinskii-Moriya interaction (DMI) and nonlocal magnetic dissipation plays a key role in the accumulation of bulk states at the boundaries. 

\end{abstract}

\maketitle

%\section{\label{sec:level1}First-level heading}
% sections are not used for PRL papers
\textit{Introduction}. For decades the application of topology in condensed matter has relied on the principle of the bulk-edge correspondence, according to which the edge states of a system, which appear under open boundary conditions, can be characterized by a topological invariant calculated on a
Brillouin zone defined under periodic boundary conditions~\cite{bernevig2013topological}. In some non-Hermitian systems, however, this fundamental correspondence has been found to be broken~\cite{kunst2018biorthogonal,gong2018topological,yao2018edge,yokomizo2019non,lee2019anatomy,song2019non,okuma2019topological,borgnia2020non,okuma2020topological,yang2020non,lee2020unraveling,lee2020ultrafast,yi2020non,jin2019bulk}. As a result, 
bulk modes can no longer be described by  Bloch’s theorem as delocalized plane waves. Instead,  a macroscopic number of bulk states localize at a boundary of the system, i.e., a phenomenon dubbed as the non-Hermitian skin effect. 

The skin effect has been extensively  investigated in one-dimensional (1$d$) asymmetric Su-Schrieffer-Heeger (SSH) models~\cite{longhi2019probing,zhu2020photonic,yao2018edge,song2019non,xu2021coexistence,lang2021dynamical,hofmann2020reciprocal}, in which the pile-up of bulk modes at one system's edge can be understood in terms of the imbalance hopping in the left and right directions. Experimentally, the skin effect has been uncovered in photonic systems and metamaterials with judiciously engineered non-Hermitian interactions, while its observation in a naturally occurring solid-state system has not yet been reported~\cite{helbig2020generalized,xiao2017observation,weidemann2020topological,zhu2020photonic,ghatak2020observation}.

Magnons, i.e., the collective excitations of magnetic systems,  are  bosonic quasiparticles whose number is not conserved and whose dynamics is intrinsically non-Hermitian~\cite{mcclarty2019non,galda2016parity,galda2019exceptional,tserkovnyak2020exceptional,jeffrey2021effect,liu2019observation}. Their fundamental properties, including  their lifetime, can be easily tuned via external fields and drives, making them promising solid-state candidates for the exploration of non-Hermitian topological phenomena~\cite{flebus2020non,zhao2020observation,zhang2019experimental,yu2020magnon}.
 In this Letter, we investigate the emergence of the skin effect in insulating two-dimensional (2$d$) magnetic systems, in which non-Hermitian terms that violate the bulk-edge correspondence  arise from intrinsic spin non-conserving interactions. 

The physics of dissipative interactions in magnetic  systems are very complex and the effective magnon lifetime stems from a variety of spin-wave decay mechanisms, e.g., magnon-magnon, magnon-electron, and magnon-phonon interactions, and magnon scattering on extrinsic impurities. Several theoretical works have addressed~\cite{streib2019magnon,balcar1971spin,wu2018magnon,rezende1978spin,berger1977simple,thompson1965intrinsic,woolsey1969theory,huber1970b} the dissipation due to one of the aforementioned mechanisms and have provided approximate expressions for the magnon relaxation time. These expressions are, however, often given in the continuum limit and can not be readily incorporated in a lattice model, which is an essential starting point for the investigation of the skin effect. 
Since a comprehensive microscopic description of the magnetic dissipative dynamics within a lattice model is a particularly challenging (and yet untackled) task, here we propose a generic phenomenological approach that can be tested against \textit{ab initio} or experimental data. Our approach serves as a general recipe of constructing an effective non-Hermitian Hamiltonian from bands broadening data. Inspired by the area law proposed by Ref.~\cite{zhang2021universal}, here we introduce an ``effective area law" that is complimentary to our phenomenological approach to magnetic dissipation and can serve as a criteria for the emergence of the skin effect in any $2d$ magnetic system.

As a concrete example of our approach, we focus on a ferromagnetic spin-orbit-coupled insulating vdW monolayer. Recent \textit{ab initio} studies have addressed the phonon-driven dissipation of the eigenmodes of a vdW magnetic system and calculated its behavior over a large portion of the first Brillouin zone~\cite{wang2021magnon}. Here, we develop a phenomenological model for the dissipative terms  that is consistent with the aforementioned \textit{ab initio} results, while respecting the symmetries of the honeycomb lattice. We find that, away from the long-wavelength limit (but below the magnetic ordering temperature), the magnetic skin effect appears, i.e., a macroscopic number of the bulk spin-wave modes accumulate at the armchair terminations of a nanoribbon. Our results show that the localization of the bulk states relies on the interference between the Dzyaloshinskii-Moriya interactions and nonlocal dissipative terms. 

\begin{figure}
\centering
{{\includegraphics[trim=0cm 0cm 0cm 0cm, clip=true,width=8.7cm, angle=0]{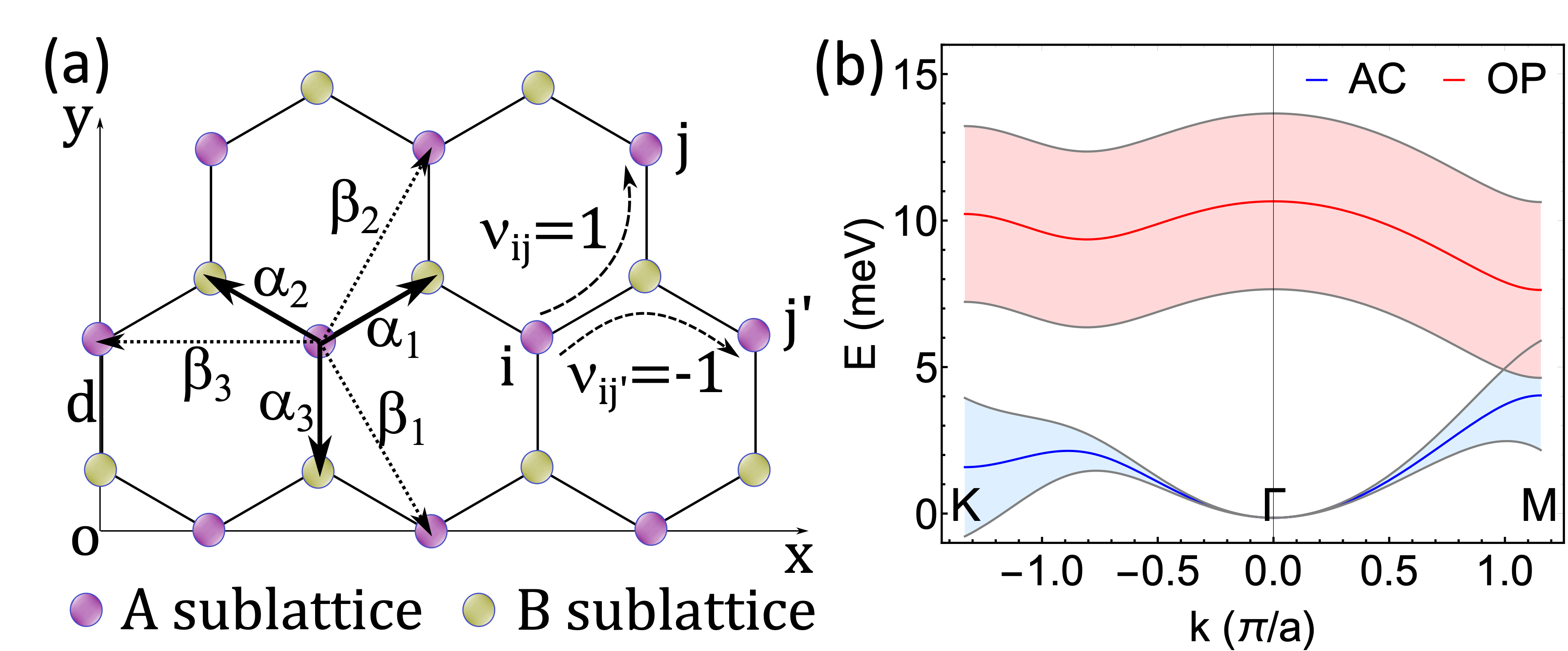}}}
\caption{(a) Ferromagnetic honeycomb lattice. $\bm{\alpha}_n$ and $\bm{\beta}_n$ are the NN and NNN bond vectors, respectively. $\nu_{ij}=1$ and $\nu_{ij'}=-1$ label the sign of counterclockwise and clockwise DMI, respectively. (b) Spin-wave dispersions with broadening along a representative path ($\Gamma-K-M$ ) in the first Brillouin zone. The blue and red lines indicate the real spectra of the acoustic (AC) and optical (OP) mode, respectively. The light blue and light red shadows represent the band broadenings $\Delta E_{\text{ac}}$ and $\Delta E_{\text{opt}}$, respectively. Here $a=\sqrt{3}d$ is the distance between A-A (B-B) sublattices, where $d$ is the NN A-B distance. }\label{Fig:1}
\end{figure}

\textit{Hermitian spin model.} We consider a vdW ferromagnetic monolayer  whose Hermitian spin dynamics is described by the Hamiltonian 
\begin{align}
\mathcal{H}=&-J\sum_{\langle i,j \rangle}\mathbf{S}_{i} \cdot\mathbf{S}_{j}-J_2\sum_{\langle\langle i,j \rangle\rangle}\mathbf{S}_{i} \cdot\mathbf{S}_{j}\nonumber\\
&-B \sum_{i}S^z_{i} 
+D\sum_{\langle \langle i, j\rangle \rangle}\nu_{i j}\hat{\mathbf{z}}\cdot(\mathbf{S}_{i}\cross\mathbf{S}_{j}),\label{h0original}
\end{align}
where  $J>0$ is the nearest-neighbor (NN) Heisenberg exchange, $J_2>0$ the next-to-nearest neighbor (NNN) exchange coupling, $B\geq 0$  the out-of-plane magnetic field, $D$ the NNN DMI strength and $\nu_{ij}=-\nu_{ji}=\pm 1$ reflects the non-reciprocity  of the DM interactions. 
Far below the magnetic ordering temperature $T_{c}$, i.e., for $T \ll T_c$, and for $D\leq (J+4J_2)/\sqrt{3}$, we can access the magnon spectrum by introducing the linearized Holstein-Primakoff transformation with respect to an uniform ground state, i.e., 
\begin{align} \label{hptransform}
S_i^+ = S_i^x + i S_i^y \approx \sqrt{2S} d_i\,, \; \; \; \; 
S_i^z = S - d_i^\dagger d_i\,,
\end{align}
where $S$ is the classical spin (in units of $\hbar$) and $d_i$ ($d^{\dagger}_{i}$) the  magnon annihilation (creation) operator at the $i$th site, which obeys the  bosonic commutation relation $[d_i, d_j^\dagger] = \delta_{ij}$.  Plugging Eq.~(\ref{hptransform}) into Eq.~\eqref{h0original} and  truncating the  Hamiltonian beyond  quadratic terms in the Holstein-Primakoff bosons, we find
\begin{align}
\mathcal{H}&=(3JS+6J_2S+B)\sum_{i}d_{i}^\dagger d_{i}-JS\sum_{\langle i,j \rangle}(d_{i}^\dagger d_{j}+\text{h.c}.)\nonumber\\
&-J_2S\sum_{\langle\langle i,j \rangle\rangle}(d_{i}^\dagger d_{j}+\text{h.c}.)-DS\sum_{\langle \langle i, j\rangle \rangle}(i\nu_{ij}d_{i}^\dagger d_{j}+\text{h.c.})\,.
\label{Hrealspace}
\end{align}
Performing a Fourier transformation while introducing the spinor $\Psi_{\bm{k}}=(a_{\bm{k}}, b_{\bm{k}})$, where $a_{\bm{k}}$ ($b_{\bm{k}}$) is the Fourier transform of the magnon annihilation operator  on the A (B) sublattice,  Eq.~(\ref{Hrealspace}) becomes
\begin{align}
\mathcal{H}= \sum_{i=0,x,y,z}\sum_{\bm{k}}\Psi_{\bm{k}}^\dagger\left(h_i\sigma_i\right)\Psi_{\bm{k}},
\label{82}
\end{align}
with
\begin{align}
h_0=&3JS+6J_2S+B-4J_2S\sum_n\cos{\bm{k}\cdot\bm{\beta}_n},\nonumber\\
h_x=&-JS\sum_{n}\cos{\bm{k}\cdot\bm{\alpha}_n},\nonumber\\
h_y=&JS\sum_{n}\sin{\bm{k}\cdot\bm{\alpha}_n},\quad h_z=2DS\sum_{n}\sin{\bm{k}\cdot\bm{\beta}_n}\,, \label{eq:hbasic}
\end{align}
where the  $\boldsymbol{\alpha}_{n}$ and  $\boldsymbol{\beta}_{n}$ (with $n=1,2,3$) are, respectively, the NN and NNN bond vectors depicted in Fig.\hyperref[Fig:1]{\ref*{Fig:1}(a)}. Here and in what follows, we omit the explicit dependence  of the function $h_i$ (for $i=x,y,z$) on the wavevector $\bm{k}$.

\textit{Non-Hermitian dissipative terms.} The magnon number is not conserved due to ubiquitous spin nonconserving interactions of magnons with the crystalline lattice~\cite{streib2019magnon,balcar1971spin,wu2018magnon,rezende1978spin,berger1977simple,thompson1965intrinsic,woolsey1969theory}. 
Several \textit{ab initio} studies have investigated the dissipation of magnetic eigenmodes driven by magnon-phonon interactions, modeled via finite-temperature random phonon fluctuations that modify the distance between neighboring spins~\cite{wang2021magnon,wang2020magnon,liu2017magnon}. 
The linewidth broadening of the acoustic, $\Delta E_{\text{ac}}$, and
optical, $\Delta E_{\text{op}}$, eigenmodes of a magnetic honeycomb lattice have been found to scale, respectively, as $\Delta E_{\text{ac}} \propto k^2$ (with $k=|\bm{k}|$) and $\Delta E_{\text{op}} \propto constant$ over a large portion of the first Brillouin zone~\cite{wang2021magnon}. While the broadening of the optical eigenmode is a constant that can be readily incorporated in the lattice Hamiltonian, we cannot include the relaxation associated with the acoustic mode in the present form $\propto k^2$ as it explicitly breaks the translational symmetry of our model~\footnote{It breaks the translational symmetry because the relaxiation in momentum space should also be periodic due to the periodicity of the Brillouin zone (BZ). For $k^2$, the relaxiation would increase to extremely large values when $\bm{k}$ is large, which is unphysical.}. 

In order to construct an effective non-Hermitian Hamiltonian that reproduces the observed band broadening, here we adopt a phenomenological approach, i.e., we include non-Hermitian terms allowed by symmetry. Specifically, we use a Fourier series (i.e.,  a complete basis) to describe a generic non-Hermitian contribution $\Delta E$ to the energy, i.e.,  $\Delta E=-i\sum_{l=0}^\infty \big[\sum_{n_1}(\zeta_{l\alpha}^1\cos{l\bm{k}\cdot\bm{\alpha}_{n_1}}+\zeta_{l\alpha}^2\sin{l\bm{k}\cdot\bm{\alpha}_{n_1}})+\sum_{n_2}(\zeta_{l\beta}^1\cos{l\bm{k}\cdot\bm{\beta}_{n_2}}+\zeta_{l\beta}^2\sin{l\bm{k}\cdot\bm{\beta}_{n_2}})+...\big]$, where ``..." represents longer crystalline vectors in higher orders. Here $n_1 (n_2)$ is the number of the crystalline vectors of the nearest (next-to-nearest) neighbors and $\zeta^{1(2)}$ is the phenomenological coefficient of each term for cosine (sine) functions in unit of energy. For purely dissipative processes, i.e., in the absence of gain, we require $\text{Im}(\Delta E) \leq0$.   For consistency, we retain terms of the same order in reciprocal vectors as the Hermitian Hamiltonian. Thus, in order to reproduce the observed broadening of the eigenmodes,  our ansatz for the imaginary part of the acoustic and optical mode eigenenergies reads as
\begin{align}
\Delta E_{\text{ac}}=&-i\chi_{11}\bigg(3-\sum_{n}\cos{\bm{k}\cdot\bm{\alpha}_n}\bigg)\nonumber \\ &-i\chi_{12}\bigg(3-\sum_{n}\cos{\bm{k}\cdot\bm{\beta}_n}\bigg)\,,\label{eq:ans2E1}\\
\Delta E_{\text{op}}=&-i\chi_{2}\label{eq:ans2E2}\,.
\end{align}

We extract the values of the parameters $\chi_{11}$, $\chi_{12}$ and $\chi_{2}$ by fitting Eqs.~(\ref{eq:ans2E1}) and~(\ref{eq:ans2E2}) to the \textit{ab initio} results of Ref.~\cite{wang2021magnon}~\footnote{See Supplementary Material for more details on the fitting.}. The band spectra and corresponding broadenings are shown in Fig.~\hyperref[Fig:1]{\ref*{Fig:1}(b)} for $\chi_{11}=1.41$ meV, $\chi_{12}=-0.415$ meV, $\chi_{2}=3$ meV,  $J=1.2$ meV, $J_2=0.02J$, $S=3/2$, $B=0.04JS=0.93$ T and $D=0.8J/\sqrt{3}$. At the $\Gamma$ point, the acoustic mode displays no broadening, in disagreement with ferromagnetic resonance measurements of  common magnetic materials. This discrepancy most likely occurs because the analysis of Ref.~\cite{wang2021magnon} neglects magnon-magnon relaxation. However, we have verified that  incorporating a constant dissipation term according to long-wavelength Landau–Lifshitz–Gilbert (LLG) phenomenology~\cite{gilbert1955lagrangian,lifshitz2013statistical,gilbert2004phenomenological,hickey2009origin} does not affect qualitatively our results.

From Eqs.~(\ref{82}),~(\ref{eq:hbasic}),~(\ref{eq:ans2E1}) and~(\ref{eq:ans2E2}),  the acoustic, $E_{\text{ac}}$, and optical, $E_{\text{op}}$, complex eigenenergies can be written as
\begin{align}
E_{\text{ac}(\text{op})}=&h_0\mp \sqrt{h_x^2+h_y^2+h_z^2}+\Delta E_{\text{ac}(\text{op})}\,.
\end{align}

The diagonal Hamiltonian $\mathcal{H}_{d}=\text{diag}(E_{\text{ac}}, E_{\text{op}})$ can be related to a non-Hermitian Hamiltonian  $\mathcal{H}_{\text{nh}}$ in the basis of the lattice operators $a_{\bm{k}}$ and $b_{\bm{k}}$ via the unitary transformation $ U \mathcal{H}_{d} U^{-1}=\mathcal{H}_{\text{nh}}$, where $U$ is a matrix composed of the eigenvectors of $\mathcal{H}$, i.e., the optical and acoustic eigenmodes of the Hermitian honeycomb lattice~\footnote{See Supplementary Material for details on this derivation.}. The Hamiltonian $\mathcal{H}_{\text{nh}}$ can be written explicitly as 
\begin{align}
\mathcal{H}_{\text{nh}}= \sum_{i=0,x,y,z}\sum_{\bm{k}}\Psi_{\bm{k}}^\dagger(\tilde{h}_i\sigma_i)\Psi_{\bm{k}},
\label{n82}
\end{align}
where $\tilde{h}_{0}=h_0+B_0$ and $\tilde{h}_{i}=h_i(1+A_0)$ for $i=x,y,z$, with
\begin{align}
A_0=\frac{\Delta E_{\text{op}}-\Delta E_{\text{ac}}}{2\sqrt{h_x^2+h_y^2+h_z^2}},\quad\quad B_0=\frac{\Delta E_{\text{op}}+ \Delta E_{\text{ac}}}{2}\,.\label{eq:A}
\end{align}

An inverse Fourier transformation yields the real-space non-Hermitian Hamiltonian as~\footnote{See Supplementary Material for the details on the inverse Fourier transformation}
\begin{align}
&\mathcal{H}_{\text{nh}}=\bigg[3JS+6J_2S+B-\frac{i}{2}(\chi_2+3\chi_{11}+3\chi_{12})\bigg]\nonumber\\
&\cross\sum_{i}(a_{i}^\dagger a_{i}+b_{i}^\dagger b_{i})\nonumber\\
+&\bigg(-J_2S+\frac{i\chi_{12}}{4}\bigg)\sum_{\langle\langle i,j \rangle\rangle}(a_{i}^\dagger a_{j}+a_{j}^\dagger a_{i}+b_{i}^\dagger b_{j}+b_{j}^\dagger b_{i})\nonumber\\
-&JS\sum_{\langle i,j\rangle}\Bigg[1-\frac{i(\chi_2-3\chi_{11}-2\chi_{12})}{2S\sqrt{3(J^2+2D^2)}}\Bigg](a_{i}^\dagger b_{j}+b_{j}^\dagger a_{i})\nonumber\\
-&DS\sum_{\langle\langle i,j \rangle\rangle}\nu_{ij}\Bigg[i+\frac{\chi_2-3\chi_{11}-2\chi_{12}}{2S\sqrt{3(J^2+2D^2)}}\Bigg]\nonumber\\
&\cross(a_{i}^\dagger a_{j}-a_{j}^\dagger a_{i}+b_{i}^\dagger b_{j}-b_{j}^\dagger b_{i})+...\label{eq:lattice-NH-all-terms}
\end{align}
where $+...$ indicates purely dissipative  higher-order-nearest-neighbor terms. By setting $\chi_{11,12,2}=0$ in Eq.~\eqref{eq:lattice-NH-all-terms}, one can recover the Hermitian Hamiltonian~(\ref{Hrealspace}). As shown by Eq.~\eqref{eq:lattice-NH-all-terms},  the non-Hermitian terms take the form of onsite dissipation terms and of nonlocal dissipative couplings, which resemble the well-known dissipative nonlocal coupling terms due to electron-mediated spin pumping~\cite{heinrich2003dynamic,tserkovnyak2003dynamic}. Here we retain only dissipative terms that have a non-dissipative counterpart; however, accounting for purely dissipative  higher-order-nearest-neighbor terms does not affect qualitatively our results~\footnote{See Supplementary Material}.

\begin{figure*}
\centering
{{\includegraphics[trim=0cm 0cm 0cm 0cm, clip=true,width=13.0cm, angle=0]{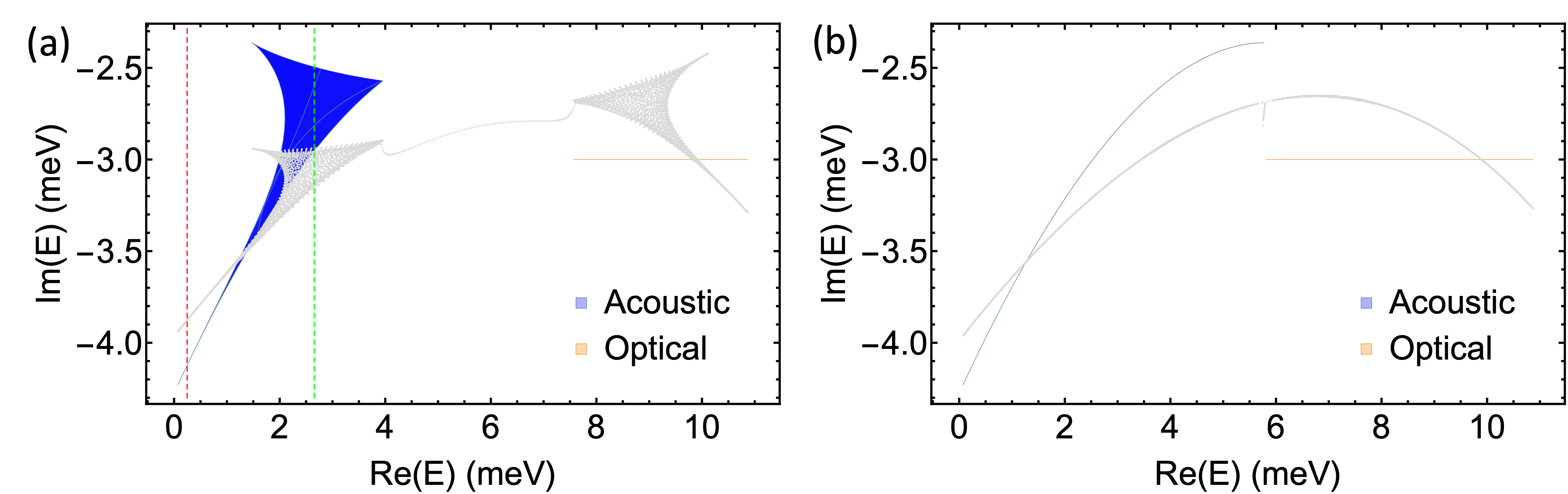}}}
{{\includegraphics[trim=0cm 0cm 0cm 0cm, clip=true,width=18.0cm, angle=0]{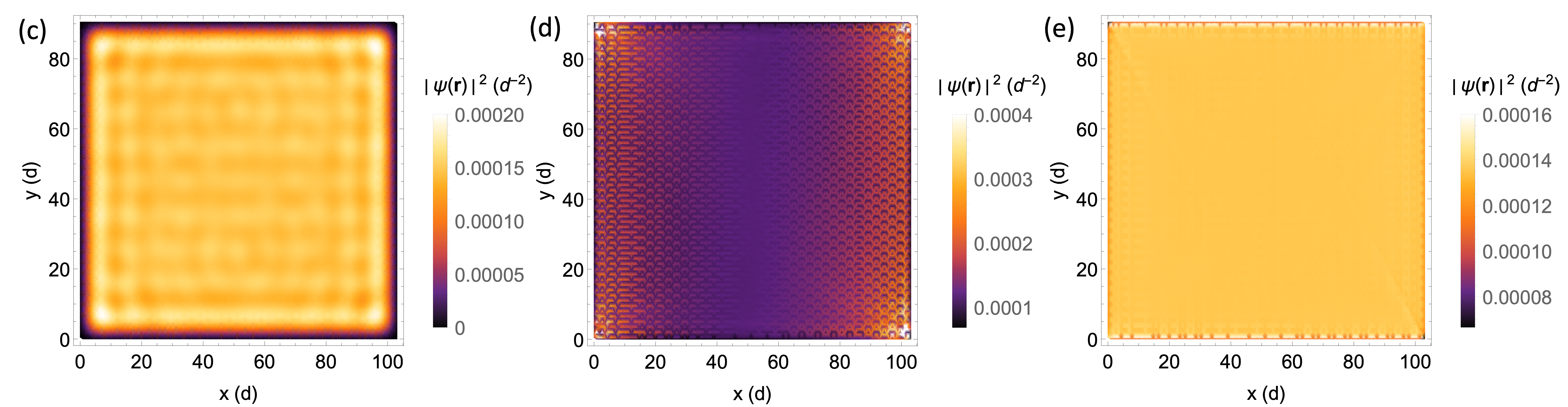}}}
\caption{(a,b)  PBC spectra of the acoustic (blue) and optical (orange) modes. The OBC eigenenergies of a nanoribbon with $60\cross 60$ lattice sites  are shown in light gray dots.  (a) For $D=0.8J/\sqrt{3}$. (b) For $D=0$.  (c-e) Spatial distribution of the density of the first $N$ right eigenstates~\eqref{psir}, (c) For $D=0.8J/\sqrt{3}$ and $N=79$, which corresponds to the energy $\text{Re}E=0.247$ meV.  This energy is indicated by  the dashed red line in Fig. 2(a), up to which the acoustic spectrum is (approximately) a line. The skin effect is absent. (d) For $D=0.8J/\sqrt{3}$ and $N=2771$, which corresponds to the energy $\text{Re}E=2.665$ meV. This energy is indicated by  the dashed green line in Fig.~2(a), around which the acoustic spectrum acquires a finite effective area. The skin effect appears. (e) For $D=0$ and $N=2771$ there is no skin effect as the acoustic spectrum reduces to the arc shown in Fig.~2(b).}\label{Fig:2}
\end{figure*}

\textit{Skin effect.} To investigate the breakdown of the bulk-edge correspondence and the emergence of the magnetic skin effect, we diagonalize the Hamiltonian~\eqref{eq:lattice-NH-all-terms} numerically under the open boundary conditions (OBC). We consider a nanoribbon  with  zigzag and chair terminations  along, respectively, the $x$- and $y$-direction. Fig.~\hyperref[Fig:2]{\ref*{Fig:2}(a)} and~\hyperref[Fig:2]{\ref*{Fig:2}(b)} show the discrepancy between the open and  periodic boundary condition (PBC) effective spectra~\footnote{The effective spectral area is obtained by neglecting the contribution proportional $\propto - \chi_{11} \sum_{j} \cos \bm{k} \cdot \boldsymbol{\alpha}_{j}$ in Eq.~(6).
More details on the effective area law we propose are discussed in the following main text and in the Supplementary Material}, which is symptomatic of a breakdown of the bulk-edge correspondence. 
From the PBC effective spectra, it is easy to see that the complex acoustic (blue) and optical (orange) bands do not cross a reference line in the complex-energy plane: thus, the system has a line gap~\cite{shen2018topological,kawabata2019symmetry}. It is worth mentioning that, while in $1d$ systems the emergence of the skin effect is ascribed to a point-gap topology~\cite{okuma2020topological}, this relation does not necessarily hold in higher-dimensional systems, in which a macroscopic accumulation of states at the boundary has been observed  in the presence of a line gap as well~\cite{lee2019hybrid}.

Analogously to its Hermitian counterpart, the inversion-symmetry-breaking DM interactions break the time-reversal symmetry of the magnon Hamiltonian, yielding a $\mathbb{Z}$ topological order (i.e., both the Hermitian and non-Hermitian Hamiltonian belong to symmetry class A). Using Fukui's algorithm~\cite{fukui2005chern}, we find $c^{\text{nB}}_{\text{ac}(\text{opt})}=\pm1$, where $c^{\text{nB}}$ is non-Bloch Chern number introduced by Ref.~\cite{yao2018non}. The corresponding topological magnon edge states can be clearly visualized in the real-energy-gapped region of the OBC spectrum shown in Fig.~\hyperref[Fig:2]{\ref*{Fig:2}(a)}. 

As a measure of the localization of the bulk eigenstates at a boundary of the ribbon, we  introduce the spatial distribution $\abs{\psi(\bm{r})}^2$ of the density of the first $N$ right eigenstates $\phi_n(\bm{r})$ of the OBC Hamiltonian~\eqref{eq:lattice-NH-all-terms}, i.e.~\cite{zhang2021universal},
\begin{align}
\abs{\psi(\bm{r})}^2=\frac{1}{N}\sum_{n=1}^{N}\abs{\phi_n(\bm{r})}^2.\label{psir}
\end{align}
Fig.~\hyperref[Fig:2]{\ref*{Fig:2}(c)} displays the spatial distribution Eq.~(\ref{psir}) of the eigenstates with energies up to  $E=0.247$ meV (corresponding to T $\sim$ 2 K with $N=79$)~\footnote{Here we include equally-weighed  contributions of magnon states up to a given energy.}, which is delocalized throughout the bulk. At higher energies, i.e., $E=2.655$ meV (T $\sim$ 30 K with $N=2771$), Fig.~\hyperref[Fig:2]{\ref*{Fig:2}(d)} shows that a macroscopic number of bulk eigenstates localizes at the corners and edges the ribbon. Thus, at high enough temperature (but below the magnetic ordering temperature $T_{c}$), the skin effect appears.

The observed temperature dependence can be understood via the relation between the skin effect and spectral shape of the PBC spectra~\cite{zhang2021universal}. In correspondence of arc or line (finite effective spectral area) in complex energy space, the mapping from momenta to energy is $2d$ to $1d$ ($2d$): for a wave impinging at the boundary there are infinite (finite) reflection channels, and an open boundary eigenstate can (can not) be described as superposition of Bloch waves, as discussed in details in Ref.~\cite{zhang2021universal}. The red and green dashed lines in Fig.~\hyperref[Fig:2]{\ref*{Fig:2}(a)} show the energies at which Fig.~\ref{Fig:2}(c) and (d) are plotted, respectively. Up to $T\sim 2$ K (red  line), the acoustic spectrum is (approximately) a line. Thus, the skin effect is not observable and bulk states behave as Bloch waves. This  result is an agreement with the conventional long-wavelength LLG treatment of magnetic dissipation~\cite{gilbert2004phenomenological}. Instead, at higher energies (green line), the spectrum acquires a finite effective area and  the skin effect appears. 

In the absence of the DM interactions, the real and imaginary part of the energy are dependent, leading to the arc-like spectrum  for both OBC and PBC displayed in Fig.~\hyperref[Fig:2]{\ref*{Fig:2}(b)}. In agreement with the area law proposed by Ref.~\cite{zhang2021universal},  the skin effect does not appear even at high temperatures, as shown in Fig.~\hyperref[Fig:2]{\ref*{Fig:2}(e)}.  Furthermore, in the absence of DMI, the Hamiltonian~(\ref{82}) becomes gapless and enters into a topologically trivial phase. The skin effect in Fig.~\hyperref[Fig:2]{\ref*{Fig:2}(d)} appears to be of the first-order-type~\cite{kawabata2020higher,zhang2021universal}, i.e., a macroscopic number of modes localizes at arbitrary edges 
due to the non-Hermitian topological properties of the Hamiltonian. As shown by Fig.~\hyperref[Fig:2]{\ref*{Fig:2}(d)}, the skin modes localized at left and right (armchair) edges, rather than at top and bottom (zigzag) terminations.  This phenomenon can be understood by calculating the point-gap winding number~\cite{okuma2020topological,zhang2021universal} while setting periodic boundary conditions only along one direction, i.e., effectively reducing the dimensionality of the system to $1d$. We find that the point-gap winding number is vanishing for PBC along the $y$ direction,  while it is non-zero for PBC along $x$ direction, which implies a localization of skin modes along the left and right edges, in agreement with Fig.~\hyperref[Fig:2]{\ref*{Fig:2}(d)}~\footnote{For more details, see Supplementary Material}.

The non-locality of the magnon-phonon driven dissipation~(\ref{eq:ans2E1}) plays also a key role. For $D\neq0$ and $\chi_{12}=0$, the skin effect does not appear even if the PBC spectrum of the acoustic eigenmode has a finite area.
When $\chi_{12}=0$, the $\bm{k}$-dependent term introducing nonlocal dissipation~(\ref{eq:ans2E1}) reduces to 
\begin{align}
\Delta E^{\text{nl}}_{\text{ac}} = \chi_{11} \sum_{n} \cos \bm{k} \cdot \boldsymbol{\alpha}_{n}\,.
\label{170}
\end{align}
By including the nonlocal dissipation~(\ref{170}) in Eq.~(\ref{82}) and performing an inverse Fourier transformation, we find that the non-Hermitian contribution~(\ref{170})  vanishes in real space due to the symmetry of the honeycomb lattice~\footnote{See the discussion about Fig.~S2 of the Supplementary Material}. 
Thus, the resulting real space Hamiltonian is equivalent to one derived by accounting only for the constant term $\propto -i \chi_{11}$ in Eq.~\eqref{eq:ans2E1}, which yields a PBC spectrum with vanishing spectral area (i.e., a line). We have verified that in this scenario the skin effect does not appear even in the high frequency regime. This suggests that the area law proposed by Ref.~\cite{zhang2021universal} should be modified in order to take into account only terms that survive upon inverse Fourier transformation and lead to a spectral area that here we call ``effective".
The nonlocal contribution $\propto \chi_{12} \sum_{n} \cos \bm{k} \cdot \boldsymbol{\beta}_{n}$, instead, is not wiped out by lattice symmetry and, in conjunction with the DM interactions, yields the skin effect.

%Ideally one would want to observe the skin effect at temperatures far below the magnetic ordering temperature, i.e., in a regime in which the magnon description is holds. 
At a given temperature $T$~\footnote{Here we consider temperatures below the magnetic ordering temperature $T_{c}$, usually $\sim 50$ K for vdW magnets}, the skin effect is maximized by  strong DM interactions, whose strength is proportional to the effective area of the PBC acoustic spectrum. A weak exchange coupling $J$ leads  to a reduced real-energy bandwidth of the acoustic magnon spectrum, which results in an amplified skin effect at a given frequency. The parameters $\chi_{11}$ and $\chi_{2}$ lead to a featureless shift of the imaginary part of the acoustic and optical eigenenergies, respectively, while $\chi_{12}$ contributes to the spectral area.
Finally, it is worth to remark that the density of right eigenstates~(\ref{psir}) plotted in Fig.~\hyperref[Fig:2]{\ref*{Fig:2}(c)} to~\hyperref[Fig:2]{\ref*{Fig:2}(e)} might be naively interpreted as a magnon local density of states  localized at an edge, which could be easily probed experimentally. However, the definition of the  physical observables of non-Hermitian systems with broken bulk-edge correspondence, which are yet relatively unexplored in two and higher dimensions, require special care since they have to be defined on a biorthogonal basis and on the real-space lattice~\cite{yi2020non,brody2013biorthogonal,kunst2018biorthogonal,curtright2007biorthogonal,chang2013biorthogonal}. We will address this problem in future investigations.

\textit{Discussion and outlook.} In this work, we have explored the emergence of the  skin effect in magnetic systems. 
We have proposed a phenomenological approach to derive  non-Hermitian Hamiltonian terms that are allowed by the symmetries of the lattice model and reproduce the band broadening observed in experimental data. We have shown that, while adopting such phenomenological approach,  the ``area law" proposed by Ref.~\cite{zhang2021universal} should be replaced by the ``effective area law"  as a criteria for the emergence of the skin effect. Our phenomenological approach, combined with the ``effective area law" ,  can be  easily applied to a wide class of dissipative magnetic systems  to identify a solid-state testbed of the skin effect that does not require \textit{ad hoc} non-Hermitian engineering and whose properties can be controlled by tuning temperature and external magnetic fields.

As an example of our method, we have focused  on a spin-orbit-coupled vdW magnet and  we have found that the skin effect appears in  at high enough temperatures  when both nonlocal dissipative terms~\footnote{We note that magnon-magnon interaction can lead as well to a $k-$dependent term to the dissipation, which could also be added to our ansatz  in a similar fashion and might lead to a stronger skin effect} and DM interactions are present. Furthermore, we have shown that the overall temperature trend of the skin effect can be understood through spectral shape of the PBC complex energy spectrum of the acoustic magnon mode. The interplay between DMI and the emergence of the skin effect should be further investigated. Future work should investigate the general physical properties that might yield the emergence of the skin effect in magnetic systems,  explore experimental protocols to probe the localization of the bulk skin modes,  and address the microscopic mechanisms underlying our phenomenological model.

\textit{Acknowledgments}. The authors thank X. Li and E. Heinrich for insightful discussions. B.F. acknowledges support of the National
Science Foundation under Grant No. NSF DMR-2144086.

\bibliographystyle{apsrev4-1}

\end{document}

\title{Supplementary Material for ``Non-Hermitian skin effect in magnetic  systems"}
\author{Kuangyin Deng}
\email{dengku@bc.edu}
\author{Benedetta Flebus}
\email{flebus@bc.edu}
\affiliation{Department of Physics, Boston College, 140 Commonwealth Avenue, Chestnut Hill, Massachusetts 02467, USA}

\maketitle

In this Supplementary Material, we discuss in detail the inclusion of non-Hermitian contributions to a two-band Hamiltonian model. We compare the band broadening adopted in the main text to the \textit{ab initio} results from Ref.~[44]. In Sec.~\ref{supsec:1}, we discuss the ineffective nonlocal dissipation terms and our effective area law. Furthermore, we provide a detailed derivation of Eq.~(11). Finally,  we discuss the role of the purely dissipative  higher-order-nearest-neighbor terms neglected in Eq.~(11). In Sec.~\ref{supsec:2}, we calculate the one-dimensional winding number in one direction by fixing the momentum in the other direction.

\section{Adding non-Hermitian contributions}\label{supsec:1}

Let us consider a 2$d$ magnetic system with two degrees of freedom, whose (Hermitian) dynamics is described by the  following Hamiltonian: 
\begin{align}
\mathcal{H}=h_0\sigma_{0}+h_x\sigma_x+h_y\sigma_y+h_z\sigma_z\,,
\label{71}
\end{align}
where the $h_i$ are functions of the 2$d$ wavevector $\bm{k}$ and $\sigma_i$ $(i=0,x,y,z)$ are the Pauli matrices. The (real) eigenvalues of Eq.~(\ref{71}) are
\begin{align}
E_{\text{op}(\text{ac})}^\text{h}=h_0\pm\sqrt{h_x^2+h_y^2+h_z^2}.
\label{76}
\end{align}
If $U$ is the corresponding eigenvector matrix, one can write $U^{-1}\mathcal{H} U=\mathcal{H}_{d}$, 
where $\mathcal{H}_{d}=\text{diag}\left( E_{\text{ac}}^\text{h},E_{\text{op}}^\text{h} \right)$ is a diagonal matrix. 
In order to account for the finite quasi-particle lifetime, we  redefine the eigenvalues~(\ref{76}) by including an imaginary term $\Delta E_{\text{op}(\text{ac})}$, i.e.,  
\begin{align}
E_{\text{op}(\text{ac})}=h_0\pm\sqrt{h_x^2+h_y^2+h_z^2}+\Delta E_{\text{op}(\text{ac})}\,,
\label{91}
\end{align}
with $\text{Re} (E_{\text{op}(\text{ac})})= E_{\text{op}(\text{ac})}^\text{h}$.
In the basis of the acoustic and optical modes with complex energies defined by Eq.~(\ref{91}), we can define a non-Hermitian Hamiltonian in the basis of its eigenvectors as $\mathcal{H}_{d, \text{nh}}=\text{diag}\left(E_{\text{ac}},E_{\text{op}}\right)$.
To obtain the non-Hermitian Hamiltonian $\mathcal{H}_{d, \text{nh}}$ in the sublattice basis, we treat the non-Hermitian contribution to the Hamiltonian as a perturbation. While this procedure might not be rigorous for the parameters used in this work, it allows us to identify the general form of the Hamiltonian in the sublattice basis. One could then use the later expression to perform directly a fitting to \textit{ab initio} or experimental data. To the leading order in the non-Hermitian perturbation, we can rewrite the non-Hermitian Hamiltonian in the  sublattice basis as 
\begin{align}
\mathcal{H}_{\text{nh}}=U\mathcal{H}_{d, \text{nh}}U^{-1}\,.
\label{90}
\end{align}

After few straightforward algebraic manipulations, we find the explicit form of Eq.~(\ref{90}) as
\begin{align}
\mathcal{H}_{\text{nh}}=\left(h_0+B_0 \right)I+h_x\left( 1+A_0\right)\sigma_x+h_y\left(1+A_0\right)\sigma_y+h_z \left(1+A_0\right)\sigma_z\,,\label{eq:Hm1-total-momentum}
\end{align}
with
\begin{align}
A_0&=\frac{\Delta E_{\text{op}}-\Delta E_{\text{ac}}}{2\sqrt{h_x^2+h_y^2+h_z^2}}\,,\quad\quad\quad
B_0=\frac{1}{2}(\Delta E_{\text{ac}}+\Delta E_{\text{op}})\,.
\label{eq:A0}
\end{align}

Up to now, our treatment is general and it is valid in the perturbative limit. Here we specialize to the Hamiltonian described by Eq.~(9) in the main text. We have included non-Hermitian terms according to Eqs.~(6) and (7) in the main text in order to reproduce the trends reported in Ref.~[44], in which the magnon-phonon driven band broadening of the optical (OP) mode is found to be a constant function of the wavevector $\bm{k}$, i.e., $\Delta E_{\text{op}} \propto \; constant$, while for  the acoustic (AC) mode we have  $\Delta E_{\text{ac}} \propto k^2$. 

\begin{figure*}
\centering
{{\includegraphics[trim=0cm 0cm 0cm 0cm, clip=true,width=15cm, angle=0]{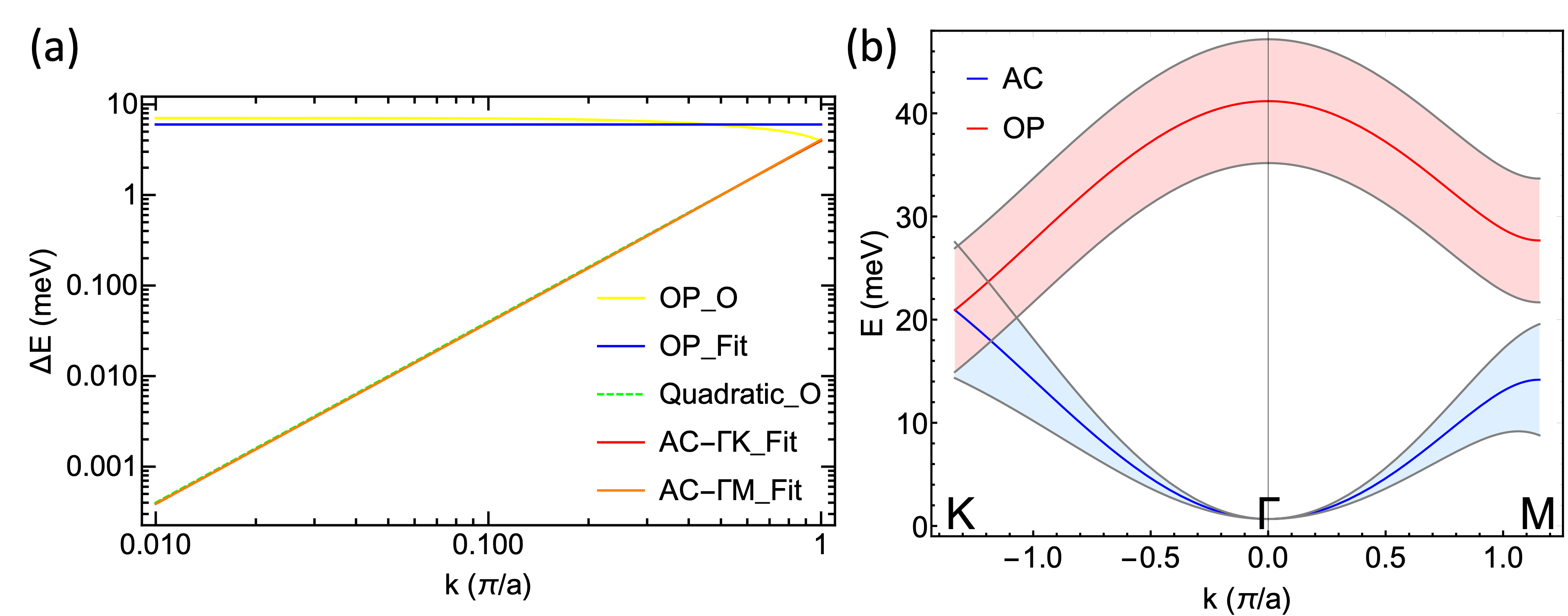}}}
\caption{(a) Band broadening $\Delta E$ on a log-log scale. The yellow (data from Ref.~[44]) and blue (fitted in our model) lines indicate the broadening of the optical mode. The dashed green line (data from Ref.~[44]), red line (fitted in our model, along the first Brillouin zone path $\Gamma-K$) and orange line (fitted in our model, along the first Brillouin zone path $\Gamma -M$) represent the broadening of the acoustic mode.  (b) Spin-wave dispersions with associated band broadening along  a representative path ($\Gamma-K-M$) in the first Brillouin zone. The blue and red lines indicate the real spectra of the acoustic and optical mode, respectively. The light blue and light red shadows represent the bands broadening of  the acoustic and optical mode obtaining by fitting our model to the parameters of Ref.~[44], respectively.}\label{Fig:S1}
\end{figure*}

It is important to note that the model of Ref.~[44] for the honeycomb lattice does not include NNN Dzyaloshinskii-Moriya $D$ and Heisenberg exchange $J_{2}$ interactions. By plugging Eqs.~(6) and (7) in the main text into Eq.~(\ref{eq:Hm1-total-momentum}) while setting $D, J_{2}=0$, $J=4.5$ meV, $S=1.5$ and $B=0.1JS$ accordingly to Ref.~[44],  we fit the acoustic and optical band broadening to the data of Ref.~[44] for temperature $T=15$ K. Our fittings, shown in Fig.~\hyperref[Fig:S1]{\ref*{Fig:S1}(a)} and~\hyperref[Fig:S1]{\ref*{Fig:S1}(b)}, leads to $\chi_{2}=6$ meV, $\chi_{11}=2.82$ meV and $\chi_{12}=-0.415$ meV. However,  these values -- obtained directly from the \textit{ab initio} calculation of Ref.~{[44]} for Cr$_2$Ge$_2$Te$_6$ --  do not yield a very strong skin effect at temperatures below the magnetic order temperature. To maximize the visibility of the skin effect, we set $J=1.2$ meV, $S=3/2$, $B=0.04JS=0.93$ T, $D=0.8J/\sqrt{3}$, $J_2=0.02J$, $\chi_{11}=1.41$ meV, $\chi_{12}=-0.415$ meV and $\chi_{2}=3$ meV. Through the main text and this supplementary material, we use these parameters if not otherwise specified. 

\subsection{Ineffective nonlocal dissipation and the  effective area law}
Here we show that -- by setting $\chi_{12}=0$ in Eq.~(6) of the main text -- the nonlocal dissipation  effectively vanishes in our real space Hamiltonian model. Inverse Fourier transforming Eq.~(\ref{eq:Hm1-total-momentum}) analytically represent a challenging task due to the denominator appearing  in Eq.~\eqref{eq:A0}, i.e.,  $\sqrt{h_x^2+h_y^2+h_z^2}$. Here, we adopt an approximation, i.e., we approximate each term in the denominator with its value averaged over the first Brillouin zone (BZ). Let us introduce
\begin{align}
\overline{h_i^2}=\frac{1}{S_k}\int{h_i^2d^2\bm{k}}\,,\label{eq:approximation2}
\end{align}
with $i=x,y,z$ and where $S_k$ is the area of the first BZ. Carrying out the integration~(\ref{eq:approximation2}) yields
\begin{align}
\overline{\Gamma}=\overline{h_x^2+h_y^2+h_z^2}=(3J^2+6D^2)S^2\,.
\label{125}
\end{align}
Plugging Eq.~(\ref{125}) into Eq.~(\ref{eq:A0}) leads to
\begin{align}
A_0&=\frac{-i}{2S\sqrt{3(J^2+2D^2)}}\bigg[\chi_{2}-\chi_{11}(3-\sum_n\cos{\bm{k}\cdot\bm{\alpha}_n})\bigg]\,, \label{222}\\
B_0&=-\frac{i}{2}\bigg[\chi_{2}+\chi_{11}(3-\sum_n\cos{\bm{k}\cdot\bm{\alpha}_n})\bigg]\,.\label{223}
\end{align}
It is important to note that the derivation above, 
which includes the approximation~(\ref{eq:approximation2}), is reported as a guidance for the reader. Since our approach is phenomenological, we could have started our discussion by taking Eqs.~(\ref{222}) and~(\ref{223}) as phenomenological ansatz, i.e., they are consistent with the symmetries of the model and, in conjunction with the term $\propto \chi_{12}$, can successfully fit the $\textit{ab initio}$ data. Thus, even in the regime in which Eq.~(\ref{eq:approximation2}) is not valid, our discussion holds.

Let us introduce the inverse Fourier transform of the lattice operators as
\begin{align}
a_{\bm{k}}&=\frac{1}{\sqrt{N_A}}\sum_{i}e^{-i\bm{k}\cdot\bm{r}_{i}}a_{i}\,,\quad\quad\quad
b_{\bm{k}}=\frac{1}{\sqrt{N_B}}\sum_{i}e^{-i\bm{k}\cdot\bm{r}_{i}}b_{i}\,,
\label{136}
\end{align}
where  $\bm{r}_{i}$ is the  position of the $i$th lattice site and $N_{A(B)}$ (with $N_{A}=N_{B}$) is the total number of A(B) lattice sites.
For simplicity, we separate the non-Hermitian Hamiltonian~(\ref{eq:Hm1-total-momentum}) in four components $\mathcal{H}_{\text{nh}}^m$, with $m=1,2,3$, as
\mathleft
\begin{align}
\mathcal{H}_{\text{nh}}=&\sum_{\bm{k}} \underbrace{(h_0+B_0)(a^\dagger_{\bm{k}}a_{\bm{k}}+b^\dagger_{\bm{k}}b_{\bm{k}})}_{m=1}+\underbrace{h_x(1+A_0)(a^\dagger_{\bm{k}}b_{\bm{k}}+b^\dagger_{\bm{k}}a_{\bm{k}})+ih_y(1+A_0)(-a^\dagger_{\bm{k}}b_{\bm{k}}+ b^\dagger_{\bm{k}}a_{\bm{k}})}_{m=2}\nonumber\\
&+\underbrace{h_z(1+A_0)(a^\dagger_{\bm{k}}a_{\bm{k}}-b^\dagger_{\bm{k}}b_{\bm{k}})}_{m=3}\,.\label{eq:Hm10}
\end{align}
Plugging Eq.~(\ref{136}) into $\mathcal{H}_{\text{nh}}^1$ leads to
\mathleft
\begin{align}
&\mathcal{H}_{\text{nh}}^1=\bigg[3JS+6J_2S+B-\frac{i}{2}(\chi_{2}+3\chi_{11})\bigg]\sum_{i}a_{i}^\dagger a_{i}-J_2S\sum_{\langle\langle i,j \rangle\rangle}(a_{i}^\dagger a_{j}+a_{j}^\dagger a_{i})+a\rightarrow b\,\nonumber\\
&+\frac{i\chi_{11}}{4N_A}\sum_{\bm{k}}\sum_{i,j}\sum_n\Big\{e^{i\bm{k}\cdot[\bm{r}_{i}-(\bm{r}_{j}+\bm{\alpha}_n)]}+e^{i\bm{k}\cdot[\bm{r}_{i}-(\bm{r}_{j}-\bm{\alpha}_n)]}\Big\}a_{i}^\dagger a_{j}\nonumber\\
&=\bigg[3JS+6J_2S+B-\frac{i}{2}(\chi_{2}+3\chi_{11})\bigg]\sum_{i}(a_{i}^\dagger a_{i}+b_{i}^\dagger b_{i})-J_2S\sum_{\langle\langle i,j \rangle\rangle}(a_{i}^\dagger a_{j}+a_{j}^\dagger a_{i}+b_{i}^\dagger b_{j}+b_{j}^\dagger b_{i})\,.\label{eq:nearestHm11}
\end{align}
Note that, in the second line, the summation over momenta $\sum_{\bm{k}}$ leads to $\delta_{\bm{r}_{i},\bm{r}_{j}+\bm{\alpha}_n}$ and $\delta_{\bm{r}_{i},\bm{r}_{j}-\bm{\alpha}_n}$. These terms vanishes because $i$ and $j$ label the same lattice pair, i.e., $i,j=$A, A (B, B), while bond vector $\bm{\alpha}_n$ connects nearest-neighbors lattice sites, i.e., A and B. Thus, Eq.~(\ref{eq:nearestHm11}) reduces to
\mathleft
\begin{align}
\mathcal{H}_{\text{nh}}^1=&\bigg[3JS+6J_2S+B-\frac{i}{2}(\chi_{2}+3\chi_{11})\bigg]\sum_{i}(a_{i}^\dagger a_{i}+b_{i}^\dagger b_{i})-J_2S\sum_{\langle\langle i,j \rangle\rangle}(a_{i}^\dagger a_{j}+a_{j}^\dagger a_{i}+b_{i}^\dagger b_{j}+b_{j}^\dagger b_{i})\,.\label{eq:nearestHnh1}
\end{align}
Plugging Eq.~(\ref{136}) into $\mathcal{H}_{\text{nh}}^2$ yields
\begin{align}
\mathcal{H}_{\text{nh}}^2=&-JS\sum_{\bm{k}}\sum_n(1+A_0)\big(\cos{\bm{k}\cdot\bm{\alpha}_n}a_{\bm{k}}^\dagger b_{\bm{k}}+\cos{\bm{k}\cdot\bm{\alpha}_n}b_{\bm{k}}^\dagger a_{\bm{k}}\nonumber\\
&+i\sin{\bm{k}\cdot\bm{\alpha}_n}a_{\bm{k}}^\dagger b_{\bm{k}}-i\sin{\bm{k}\cdot\bm{\alpha}_n}b_{\bm{k}}^\dagger a_{\bm{k}}\big)\nonumber\\
=&-JS\sum_{\bm{k}}\Bigg\{1-\frac{i}{2S\sqrt{3(J^2+2D^2)}}\bigg[\chi_{2}-\chi_{11}(3-\sum_{m}\cos{\bm{k}\cdot\bm{\alpha}_{m}})\bigg]\Bigg\}\nonumber\\
&\cross\sum_n(e^{i\bm{k}\cdot\bm{\alpha}_n}a_{\bm{k}}^\dagger b_{\bm{k}}+e^{-i\bm{k}\cdot\bm{\alpha}_n}b_{\bm{k}}^\dagger a_{\bm{k}})\nonumber\\
=&-JS\sum_{\langle i,j \rangle }\Bigg[1-\frac{i(\chi_{2}-3\chi_{11})}{2S\sqrt{3(J^2+2D^2)}}\Bigg](a_{i}^\dagger b_{j}+b_{j}^\dagger a_{i})\,.\label{eq:nearestHnh2}
\end{align}
Similarly to Eq.~\eqref{eq:nearestHm11}, the contributions from terms $\propto \sum_{m}\cos{\bm{k}\cdot\bm{\alpha}_{m}}$ vanish since they lead to terms $\propto \delta_{\bm{r}_{i},\bm{r}_{j}-\bm{\alpha}_{m}-\bm{\alpha}_{n}}$, while $i$ and $j$ connect the nearest-neighbor sites A-B or B-A.  Similarly, for the Hamiltonian term $\mathcal{H}_{\text{nh}}^3$ the contributions due to the term $\sum_{m}\cos{\bm{k}\cdot\bm{\alpha}_{m}}$ do not survive either. We obtain
\mathleft
\begin{align}
\mathcal{H}_{\text{nh}}^3=&-DS\sum_{\langle\langle i,j \rangle\rangle}\Bigg[1-\frac{i(\chi_{2}-3\chi_{11})}{2S\sqrt{3(J^2+2D^2)}}\Bigg](i\nu_{ij}a_{i}^\dagger a_{j}-i\nu_{ij}a_{j}^\dagger a_{i}+i\nu_{ij}b_{i}^\dagger b_{j}-i\nu_{ij}b_{j}^\dagger b_{i})\,.\label{eq:nearestHnh3}
\end{align}
Summing Eq.~\eqref{eq:nearestHnh1},~\eqref{eq:nearestHnh2} and~\eqref{eq:nearestHnh3} leads to
\mathleft
\begin{align}
\mathcal{H}_{\text{nh}}^\text{R}&=\bigg[3JS+6J_2S+B-\frac{i}{2}(\chi_{2}+3\chi_{11})\bigg]\sum_{i}(a_{i}^\dagger a_{i}+b_{i}^\dagger b_{i})\nonumber\\
&-JS\sum_{\langle i,j \rangle }\Bigg[1-\frac{i(\chi_{2}-3\chi_{11})}{2S\sqrt{3(J^2+2D^2)}}\Bigg](a_{i}^\dagger b_{j}+b_{j}^\dagger a_{i})-J_2S\sum_{\langle\langle i,j \rangle\rangle}(a_{i}^\dagger a_{j}+a_{j}^\dagger a_{i}+b_{i}^\dagger b_{j}+b_{j}^\dagger b_{i})\nonumber\\
&-DS\sum_{\langle\langle i,j \rangle\rangle}\Bigg[1-\frac{i(\chi_{2}-3\chi_{11})}{2S\sqrt{3(J^2+2D^2)}}\Bigg](i\nu_{ij}a_{i}^\dagger a_{j}-i\nu_{ij}a_{j}^\dagger a_{i}+i\nu_{ij}b_{i}^\dagger b_{j}-i\nu_{ij}b_{j}^\dagger b_{i})\,.
\end{align}
Thus, in the sublattice basis, the contribution from terms  $\propto \sum_{n}\cos{\bm{k}\cdot\bm{\alpha}_{n}}$ in Eq.~(\ref{222}) and ~(\ref{223}) vanishes. Thus, the resulting Hamiltonian  is equivalent to the one that can be derived from the following non-Hermitian terms
\mathcenter
\begin{align}
A_0&=\frac{-i(\chi_{2}-3\chi_{11})}{2S\sqrt{3(J^2+2D^2)}}\,, \quad\quad\quad
B_0=-\frac{i}{2}(\chi_{2}+3\chi_{11})\,.\label{eq:nearest-B}
\end{align}

\begin{figure}
\centering
{{\includegraphics[trim=0cm 0cm 0cm 0cm, clip=true,width=16.0cm, angle=0]{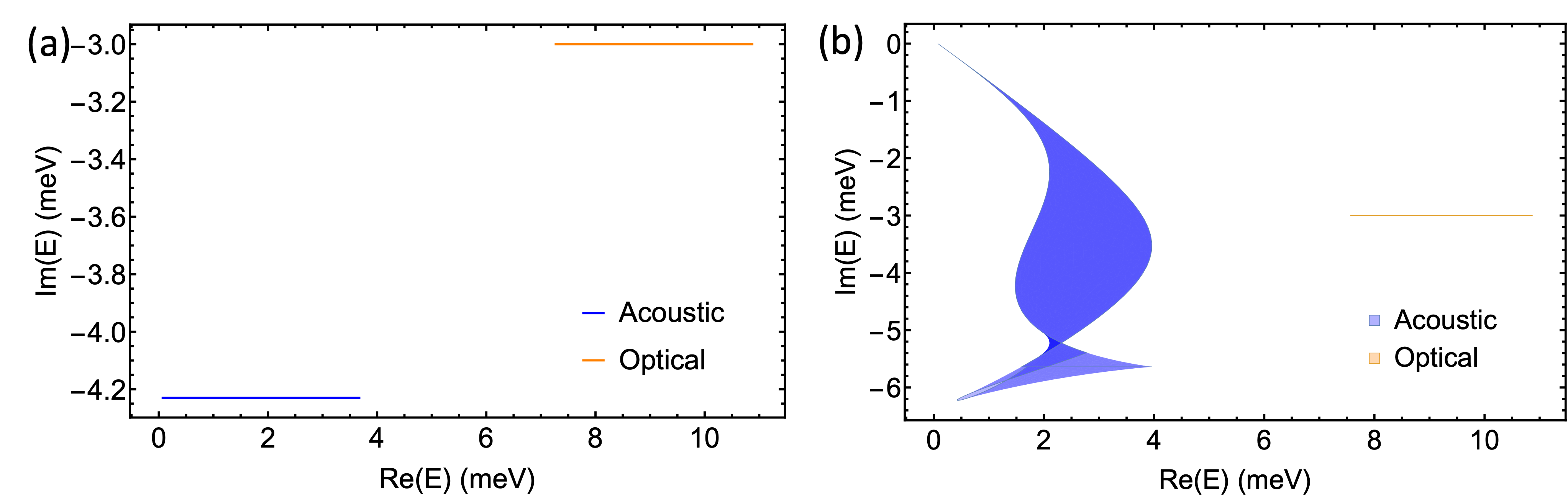}}}
\caption{(a) Effective complex energy spectra of the acoustic (blue) and optical (orange) modes for Eq.~(\ref{eq:nearest-B}). The acoustic spectrum reduces to a line. (b) The whole acoustic (blue) and optical (orange) spectra in complex plane according to Eqs.~(\ref{222}) and~(\ref{223}) while setting $\chi_{12}=0$. The acoustic spectra (blue) has a finite area spectral that is effectively equivalent to the blue line displayed in (a) upon inverse Fourier transformation.}\label{Fig:S2}
\end{figure}

The complex energy spectra associated with the non-Hermitian contribution in Eq.~(\ref{eq:nearest-B}) are two lines, as shown in Fig.~\hyperref[Fig:S2]{\ref*{Fig:S2}(a)}. Thus, the skin effect can not be observed at any temperature, as we have numerically verified. Fig.~\hyperref[Fig:S2]{\ref*{Fig:S2}(b)} shows that the spectral area of the complex spectrum corresponding to Eqs.~(\ref{222}) and~(\ref{223}) is  finite, which contradicts the spectral area law proposed by Ref.~[43]. Our results suggest that the spectral area law should be revised by accounting only for an ``effective area" due to the non-Hermitian contribution that survive in real-space representation. Following this reasoning, in Fig.~2(a) and~2(b) of the main text, we display the spectral area associated with Eqs.~(6) and (7) while subtracting the area due to the ineffective contribution $\propto -i\chi_{12} \sum_n\cos{\bm{k}\cdot\bm{\alpha}_n}$.

\subsection{Effective nonlocal dissipation}
In this section, we derive the Hamiltonian in Eq.~(11) of the main text, while dropping the ineffective nonlocal dissipation term from Eqs.~(6) and (7).
Invoking the approximation~(\ref{125}), we can rewrite Eqs.~(6) and~(7) as  
\begin{align}
\Delta E_{\text{ac}}&=-3i\chi_{11}-i\chi_{12}\bigg(3-\sum_n\cos{\bm{k}\cdot\bm{\beta}_n}\bigg)\,,\label{eq:ans2E12}\\
\Delta E_{\text{op}}&=-i\chi_{2}\,.\label{eq:ans2E22}
\end{align}
We can generally rewrite the non-Hermitian Hamiltonian as Eq.~(\ref{eq:Hm10}). Inverse Fourier transforming $\mathcal{H}_{\text{nh}}^1$ leads to
\mathleft
\begin{align}
\mathcal{H}_{\text{nh}}^1
=&\bigg[3JS+6J_2S+B-\frac{i}{2}(\chi_{2}+3\chi_{11}+3\chi_{12})\bigg]\sum_{i}a_{i}^\dagger a_{i} -J_2S\sum_{\langle\langle i,j \rangle\rangle}(a_{i}^\dagger a_{j}+a_{j}^\dagger a_{i})+a\rightarrow b\nonumber\\
&+\frac{i\chi_{12}}{4N_A}\sum_{\bm{k}}\sum_{i,j}\sum_n\Big\{e^{i\bm{k}\cdot[\bm{r}_{i}-(\bm{r}_{j}+\bm{\beta}_n)]}+e^{i\bm{k}\cdot[\bm{r}_{i}-(\bm{r}_{j}-\bm{\beta}_n)]}\Big\}a_{i}^\dagger a_{j}\nonumber\\
=&\bigg[3JS+6J_2S+B-\frac{i}{2}(\chi_{2}+3\chi_{11}+3\chi_{12})\bigg]\sum_{i}(a_{i}^\dagger a_{i}+b_{i}^\dagger b_{i})\nonumber\\
&+\bigg(-J_2S+\frac{i\chi_{12}}{4}\bigg)\sum_{\langle\langle i,j \rangle\rangle}(a_{i}^\dagger a_{j}+a_{j}^\dagger a_{i}+b_{i}^\dagger b_{j}+b_{j}^\dagger b_{i})\,.\label{eq:allHnh1}
\end{align}
For the term $\mathcal{H}_{\text{nh}}^2$, we have 
\mathleft
\begin{align}
\mathcal{H}_{\text{nh}}^2=&-JS\sum_{\bm{k}}\sum_n(1+A_0)\big(\cos{\bm{k}\cdot\bm{\alpha}_n}a_{\bm{k}}^\dagger b_{\bm{k}}+\cos{\bm{k}\cdot\bm{\alpha}_n}b_{\bm{k}}^\dagger a_{\bm{k}}\nonumber\\
&+i\sin{\bm{k}\cdot\bm{\alpha}_n}a_{\bm{k}}^\dagger b_{\bm{k}}-i\sin{\bm{k}\cdot\bm{\alpha}_n}b_{\bm{k}}^\dagger a_{\bm{k}}\big)\nonumber\\
=&-JS\sum_{\langle i,j \rangle }\Bigg[1-\frac{i(\chi_{2}-3\chi_{11}-3\chi_{12})}{2S\sqrt{3(J^2+2D^2)}}\Bigg](a_{i}^\dagger b_{j}+b_{j}^\dagger a_{i})\nonumber\\
&+JS\frac{i\chi_{12}}{2S\sqrt{3(J^2+2D^2)}}\sum_{n,m}\frac{1}{2N_A}\sum_{i,j}\sum_{\bm{k}}(e^{i\bm{k}\cdot\bm{\beta}_{m}}+e^{-i\bm{k}\cdot\bm{\beta}_{m}})\nonumber\\
&\cross\Big[e^{i\bm{k}\cdot(\bm{r}_{i}-\bm{r}_{j}+\bm{\alpha}_n)}a_{i}^\dagger b_{j}+e^{i\bm{k}\cdot(-\bm{r}_{i}+\bm{r}_{j}-\bm{\alpha}_n)}b_{j}^\dagger a_{i}\Big]\nonumber
\end{align}
\mathleft
\begin{align}
\quad\quad=&-JS\sum_{\langle i,j \rangle }\Bigg[1-\frac{i(\chi_{2}-3\chi_{11}-3\chi_{12})}{2S\sqrt{3(J^2+2D^2)}}\Bigg](a_{i}^\dagger b_{j}+b_{j}^\dagger a_{i})\nonumber\\
&+JS\frac{i\chi_{12}}{4S\sqrt{3(J^2+2D^2)}}\sum_{n,m}\sum_{i,j}\Big\{\big[\delta(\bm{r}_{i},\bm{r}_{j}-\bm{\alpha}_n-\bm{\beta}_{m})+\delta(\bm{r}_{i},\bm{r}_{j}-\bm{\alpha}_n+\bm{\beta}_{m})\big]a_{i}^\dagger b_{j}\nonumber\\
&+\big[\delta(\bm{r}_{i},\bm{r}_{j}-\bm{\alpha}_n+\bm{\beta}_{m})+\delta(\bm{r}_{i},\bm{r}_{j}-\bm{\alpha}_n-\bm{\beta}_{m})\big]b_{j}^\dagger a_{i}\Big\}\nonumber\\
=&-JS\sum_{\langle i,j \rangle }\Bigg[1-\frac{i(\chi_{2}-3\chi_{11}-2\chi_{12})}{2S\sqrt{3(J^2+2D^2)}}\Bigg](a_{i}^\dagger b_{j}+b_{j}^\dagger a_{i})\nonumber\\
&+JS\frac{i\chi_{12}}{2S\sqrt{3(J^2+2D^2)}}\bigg(\sum_{\langle i,j \rangle ^3}+\frac{1}{2}\sum_{\langle i,j \rangle ^4}\bigg)(a_{i}^\dagger b_{j}+b_{j}^\dagger a_{i})\,,
\label{eq:allHnh2}
\end{align}
where we denote $\langle i,j \rangle ^n$ the $n$th neighbor. Finally, inverse Fourier transforming $\mathcal{H}_{\text{nh}}^3$ yields
\mathleft
\begin{align}
\mathcal{H}_{\text{nh}}^3=&\sum_{\bm{k}}h_z(1+A_0)(a^\dagger_{\bm{k}}a_{\bm{k}}-b^\dagger_{\bm{k}}b_{\bm{k}})\nonumber\\
=&2DS\sum_{\bm{k}}\sum_n\sin{\bm{k}\cdot\bm{\beta}_n}\Bigg\{1-\frac{i\big[\chi_{2}-3\chi_{11}-\chi_{12}(3-\sum_{m}\cos{\bm{k}\cdot\bm{\beta}_{m}})\big]}{2S\sqrt{3(J^2+2D^2)}}\Bigg\}(a^\dagger_{\bm{k}}a_{\bm{k}}-b^\dagger_{\bm{k}}b_{\bm{k}})\nonumber\\
=&-DS\sum_{\langle\langle i,j \rangle\rangle}\Bigg[1-\frac{i(\chi_{2}-3\chi_{11}-3\chi_{12})}{2S\sqrt{3(J^2+2D^2)}}\Bigg]i\nu_{ij}(a_{i}^\dagger a_{j}-a_{j}^\dagger a_{i}+b_{i}^\dagger b_{j}-b_{j}^\dagger b_{i})\nonumber\\
&+DS\sum_{\bm{k}}\sum_n(e^{i\bm{k}\cdot\bm{\beta}_n}-e^{-i\bm{k}\cdot\bm{\beta}_n})(-i)\frac{-i\chi_{12}}{2S\sqrt{3(J^2+2D^2)}}\frac{1}{2N_A}\sum_{m}(e^{i\bm{k}\cdot\bm{\beta}_{m}}+e^{-i\bm{k}\cdot\bm{\beta}_{m}})\nonumber\\
&\cross\sum_{i,j}e^{i\bm{k}\cdot(\bm{r}_{i}-\bm{r}_{j})}(a_{i}^\dagger a_{j}-b_{i}^\dagger b_{j})\nonumber
\end{align}
\begin{figure*}
\centering
{{\includegraphics[trim=0cm 0cm 0cm 0cm, clip=true,width=12.0cm, angle=0]{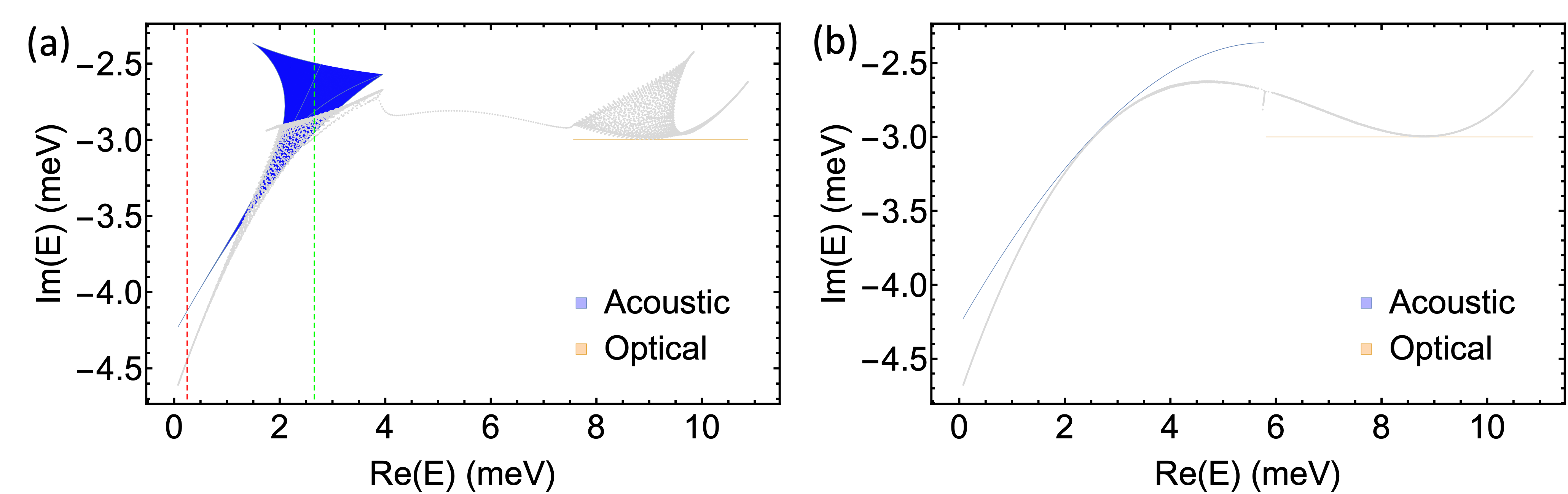}}}
{{\includegraphics[trim=0cm 0cm 0cm 0cm, clip=true,width=16.5cm, angle=0]{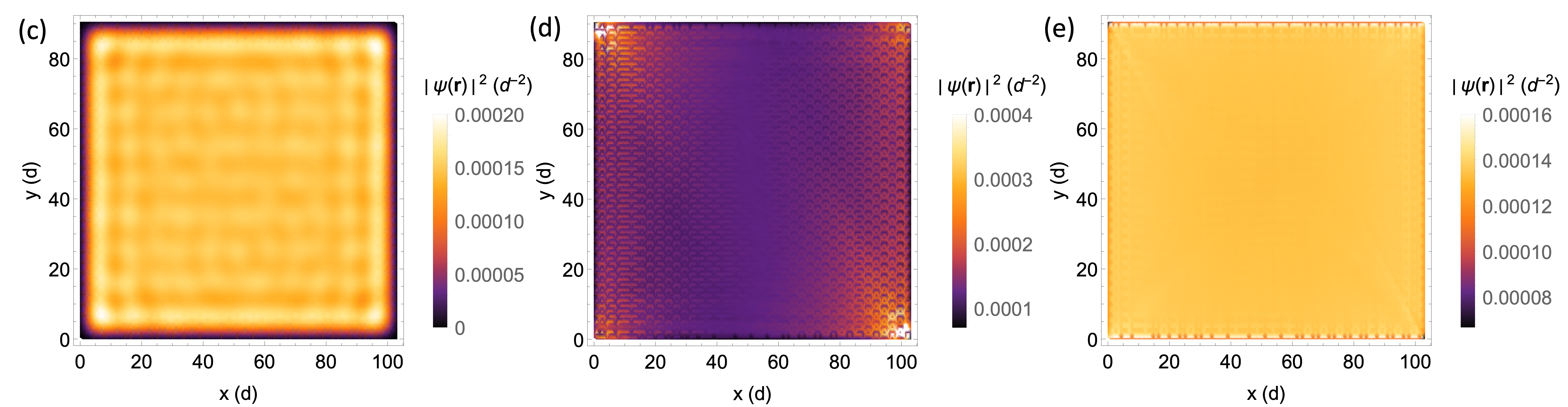}}}
\caption{(a, b) PBC spectra of the acoustic (blue) and optical (orange) modes. The OBC eigenenergies calculated from Eq. S24 are shown in light gray dots.  (c-e) Spatial distribution of the density of the first $N$ right eigenstates calculated from Eq. S24. All  parameters are analogous to ones used in Fig. 2 of the main text.}\label{Fig:S3}
\end{figure*}
\begin{align}
\quad\quad=&-DS\sum_{\langle\langle i,j \rangle\rangle}\Bigg[1-\frac{i(\chi_{2}-3\chi_{11}-3\chi_{12})}{2S\sqrt{3(J^2+2D^2)}}\Bigg]i\nu_{ij}(a_{i}^\dagger a_{j}-a_{j}^\dagger a_{i}+b_{i}^\dagger b_{j}-b_{j}^\dagger b_{i})\nonumber\\
&-DS\frac{\chi_{12}}{2S\sqrt{3(J^2+2D^2)}}\frac{1}{2}\sum_{n,m,i,j}\Big[\delta(\bm{r}_{i},\bm{r}_{j}-\bm{\beta}_n-\bm{\beta}_{m})+\delta(\bm{r}_{i},\bm{r}_{j}-\bm{\beta}_n+\bm{\beta}_{m})\nonumber\\
&-\delta(\bm{r}_{i},\bm{r}_{j}+\bm{\beta}_n-\bm{\beta}_{m})-\delta(\bm{r}_{i},\bm{r}_{j}+\bm{\beta}_n+\bm{\beta}_{m})\Big](a_{i}^\dagger a_{j}-b_{i}^\dagger b_{j})\nonumber\\
=&-DS\sum_{\langle\langle i,j \rangle\rangle}\nu_{ij}\Bigg[i+\frac{\chi_{2}-3\chi_{11}-2\chi_{12}}{2S\sqrt{3(J^2+2D^2)}}\Bigg](a_{i}^\dagger a_{j}-a_{j}^\dagger a_{i}+b_{i}^\dagger b_{j}-b_{j}^\dagger b_{i})\nonumber\\
&-DS\frac{\chi_{12}}{2S\sqrt{3(J^2+2D^2)}}\sum_{\langle i,j \rangle ^6}\frac{\mu_{ij}}{2}(-a_{i}^\dagger a_{j}+a_{j}^\dagger a_{i}-b_{i}^\dagger b_{j}+b_{j}^\dagger b_{i})\,,\label{eq:allHnh3}
\end{align}
where $\mu_{i,j}$ connects two sites with distance of $2\bm{\beta}_n$ Denoting $\bm{r}_{i_M}=(\bm{r}_{i}+\bm{r}_{j})/2$, one can set $\mu_{i,j}=\nu_{i_M,j}$.  Summing together Eqs.~\eqref{eq:allHnh1},~\eqref{eq:allHnh2} and~\eqref{eq:allHnh3} we find
\begin{align}
\mathcal{H}_{\text{nh}}^\text{R}
=&\bigg[3JS+6J_2S+B-\frac{i}{2}(\chi_{2}+3\chi_{11}+3\chi_{12})\bigg]\sum_{i}(a_{i}^\dagger a_{i}+b_{i}^\dagger b_{i})\nonumber\\
&+\bigg(-J_2S+\frac{i\chi_{12}}{4}\bigg)\sum_{\langle\langle i,j \rangle\rangle}(a_{i}^\dagger a_{j}+a_{j}^\dagger a_{i}+b_{i}^\dagger b_{j}+b_{j}^\dagger b_{i})\nonumber\\
&-JS\sum_{\langle i,j \rangle }\Bigg[1-\frac{i(\chi_{2}-3\chi_{11}-2\chi_{12})}{2S\sqrt{3(J^2+2D^2)}}\Bigg](a_{i}^\dagger b_{j}+b_{j}^\dagger a_{i})\nonumber\\
&-DS\sum_{\langle\langle i,j \rangle\rangle}\nu_{ij}\Bigg[i+\frac{\chi_{2}-3\chi_{11}-2\chi_{12}}{2S\sqrt{3(J^2+2D^2)}}\Bigg](a_{i}^\dagger a_{j}-a_{j}^\dagger a_{i}+b_{i}^\dagger b_{j}-b_{j}^\dagger b_{i})\nonumber\\
&+JS\frac{i\chi_{12}}{2S\sqrt{3(J^2+2D^2)}}\bigg(\sum_{\langle i,j \rangle ^3}+\frac{1}{2}\sum_{\langle i,j \rangle ^4}\bigg)(a_{i}^\dagger b_{j}+b_{j}^\dagger a_{i})\nonumber\\
&-DS\frac{\chi_{12}}{2S\sqrt{3(J^2+2D^2)}}\sum_{\langle i,j \rangle ^6}\frac{\mu_{ij}}{2}(-a_{i}^\dagger a_{j}+a_{j}^\dagger a_{i}-b_{i}^\dagger b_{j}+b_{j}^\dagger b_{i})\,,
\label{final Hamiltonian}
\end{align}
where the last two lines are  higher-order-nearest-neighbor terms  that do not appear explicitly  in Eq.~(11) of the main text. In the main text, we neglect them since, similarly to the electron-mediated spin pumping terms, one would naturally expect a dissipative interaction to have a nondissipative counterpart. This assumption does not affect qualitatively our results: as shown by Fig.~\hyperref[Fig:S3]{\ref*{Fig:S3}}, the complex energy spectra, and  the dependence of the skin effect on DMI and temperature  are comparable with the ones shown in Fig.~2 of the main text. 

A finally remark on our approximations [see Eqs.~(\ref{90}) and~(\ref{125})] is in place. While these approximations might not hold for certain parameters, they allow  us to derive the structure of  the non-Hermitian  Hamiltonian~(\ref{final Hamiltonian}) in sublattice basis starting from our phenomenological ansatz, i.e. Eqs.~(6) and~(7). Alternatively, one could avoid invoking these approximations by starting from a Hamiltonian with the structure of Eq.~(\ref{final Hamiltonian}) and use fitting parameters to reproduce the results of Ref.~[44].

\section{One-dimensional winding number}\label{supsec:2}

In Fig. 2(d) of the main text, we observe a first-order skin effect accumulating magnons at the armchair edges (left and right) rather than at the zigzag terminations (up and down). Here we show that this anisotropic localization can be understood in terms of one-dimensional winding numbers in the $x$ and $y$ directions.

\begin{figure*}
\centering
{{\includegraphics[trim=0cm 0cm 0cm 0cm, clip=true,width=17cm, angle=0]{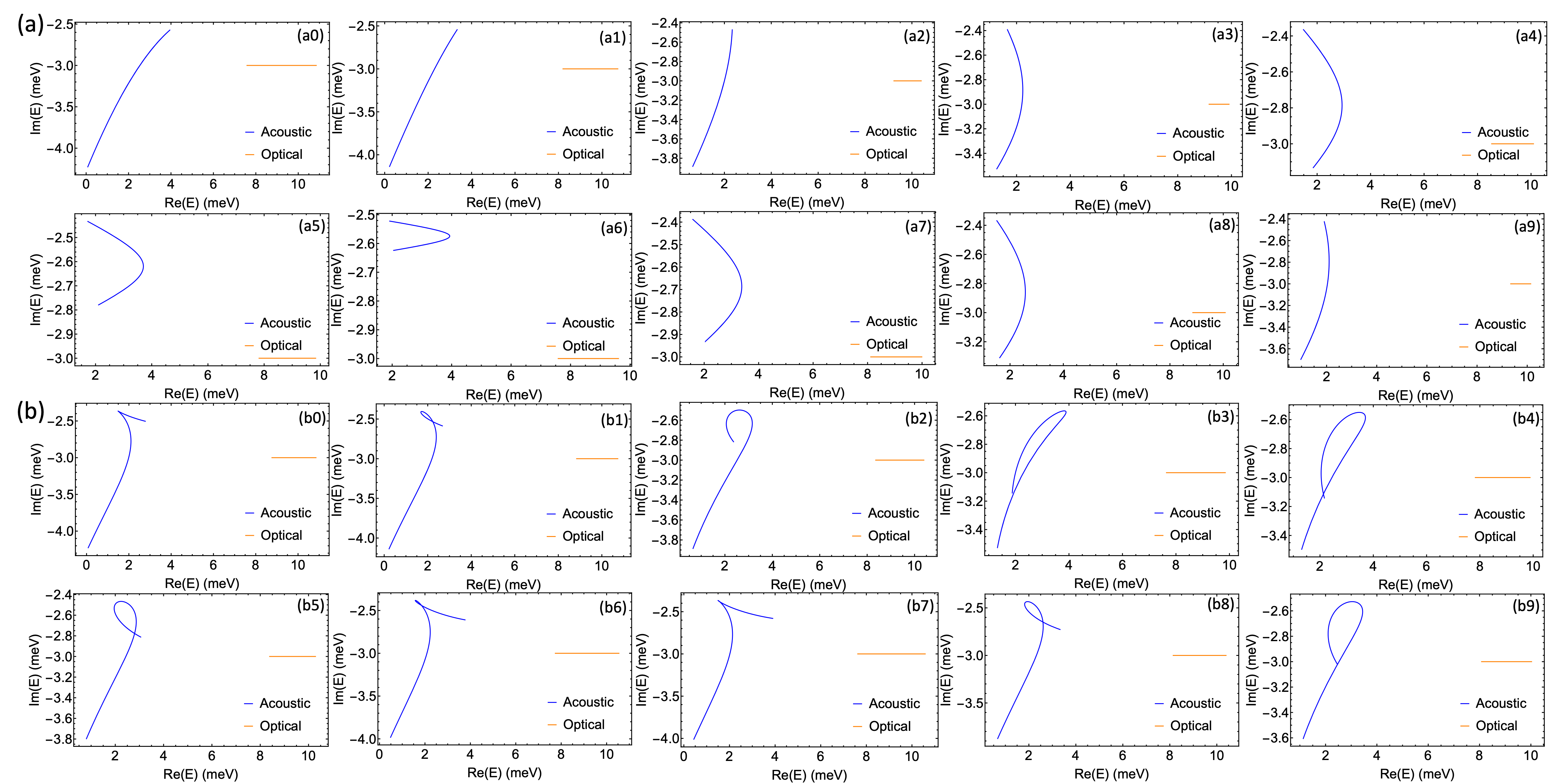}}}
\caption{{Complex energy spectra obtain by setting $k_x$ and $k_y$, respectively, to a fixed value. Here we set the lattice constant  $d=1$. (a0) to (a9) correspond to $k_x=\frac{n \pi}{10}$ and (b0) to (b9) correspond to $k_y=\frac{n \pi}{10}$, with $n=0, ..., 9$. The winding number  shown in (a) is always 0, while in (b) we find a non-zero winding number  for most values of $k_y$. This confirms the anisotropic first-order skin effect shown in Fig. 2(d) of the main text.}}\label{Fig:S4}
\end{figure*}

While the emergence of the skin effect in higher dimensions is still relatively unexplored, in one-dimensional ($1d$) systems it is clear that  the skin effect appears in correspondence of a non-zero winding number associated with a point-gap topology (See Refs.~[10, 43] in the main text).  
We have investigated the winding numbers  by fixing $k_x$ or $k_y$, i.e., by allowing for PBC only in the $y$ and $x$ direction respectively, thus effectively reducing the dimensionality of our problem to $1d$.  The formal calculation of the $1d$ winding number is equivalent to graphically check if the resulting spectrum winds around a reference energy point in the complex energy plane.  We find that the point-gap winding number is zero  for fixed $k_x$, as shown in Fig.~\hyperref[Fig:S4]{\ref*{Fig:S4}(a)}, while it is non-zero  for most of the fixed values of $k_y$, as shown in Fig.~\hyperref[Fig:S4]{\ref*{Fig:S4}(b)}. The later finding suggests that the macroscopic accumulation of states should occur along the left and right edge, instead of the  top and bottom edges, in agreement with our results.

